\documentclass[conference]{IEEEtran}
\IEEEoverridecommandlockouts

\usepackage{cite}
\usepackage{amsmath,amssymb,color,graphicx,caption,subcaption,comment,tikz,url,latexsym} 
\usepackage{algorithmic}
\usepackage{graphicx}
\usepackage{textcomp}
\usepackage{pgfplots}
\usepackage{xcolor}
\usepackage{url}
\usepackage{multirow}
\usepackage{mathtools}
\usepackage[left=0.62in,right=0.62in,top=0.6in,bottom=0.6in]{geometry}

\usepgfplotslibrary{units}
\usetikzlibrary{spy,backgrounds}
\usepackage{pgfplotstable}

\def\BibTeX{{\rm B\kern-.05em{\sc i\kern-.025em b}\kern-.08em
		T\kern-.1667em\lower.7ex\hbox{E}\kern-.125emX}}

\usepackage{cite}
\usepackage{amsmath,amssymb,color,graphicx,caption,subcaption,comment,tikz,url,latexsym} 
\usepackage{graphicx}
\usepackage{textcomp}
\usepackage{pgfplots}
\usepackage{xcolor}
\usepackage{url}
\usepackage{multirow}
\usepackage{bbm}
\usepackage{dsfont}
\usepackage{mathtools}
\usepgfplotslibrary{units}
\usetikzlibrary{spy,backgrounds}
\usepackage{pgfplotstable}

\def\BibTeX{{\rm B\kern-.05em{\sc i\kern-.025em b}\kern-.08em
		T\kern-.1667em\lower.7ex\hbox{E}\kern-.125emX}}

\newtheorem{theorem}{Theorem} 

\newtheorem{definition}{Definition}
\newtheorem{lemma}{Lemma}

\usepackage{bbm}
\newcommand{\M}{\mathsf{M}}
\newcommand{\m}{\mathsf{m}}
\newcommand{\len}{\textrm{len}}

\newcommand{\E}{\mathbb{E}}

\newcommand{\D}{\mathcal{D}}

\newcommand{\oi}{\text{\o}}

\begin{document}
	\interdisplaylinepenalty=0
	\title{Strong Converses using Change of Measure and Asymptotic Markov Chains\\
	}
	\author{\IEEEauthorblockN{Mustapha Hamad}
		\IEEEauthorblockA{\textit{LTCI, Telecom Paris, IP Paris} \\
			91120 Palaiseau, France\\
			mustapha.hamad@telecom-paris.fr}
		\and
		\IEEEauthorblockN{Mich\`ele Wigger}
		\IEEEauthorblockA{\textit{LTCI, Telecom Paris, IP Paris} \\
			91120 Palaiseau, France\\
			michele.wigger@telecom-paris.fr}
		\and
		\IEEEauthorblockN{Mireille Sarkiss}
		\IEEEauthorblockA{\textit{SAMOVAR, Telecom SudParis, IP Paris} \\
			91011 Evry, France\\
			mireille.sarkiss@telecom-sudparis.eu}
	}
	\allowdisplaybreaks[4]
	\sloppy
	\maketitle

\begin{abstract}
The main contribution of this paper is a strong converse result for  $K$-hop distributed hypothesis testing against independence with multiple (intermediate) decision centers under a Markov condition. Our result shows that the set of type-II error exponents that can simultaneously be achieved at all the terminals  does not depend on the maximum permissible  type-I error probabilities. Our strong converse proof is based on a change of measure argument and on the asymptotic proof of specific  Markov chains. This proof  method can also be used for other converse proofs, and  is appealing because it does not require resorting to variational characterizations or blowing-up methods as in previous related proofs.
\end{abstract}
\section{Introduction}

Strong converse results have a rich history in information theory. They refer to  proofs showing  that the  fundamental performance limit (such as minimum compression rate or capacity) of  a  specific system does not depend on its asymptotically allowed error (or excess) probability (as long it is not $1$). For example, Wolfowitz' strong converse \cite{Wolfowith78}  established that the  capacity of a discrete-memoryless channel remains unchanged  when positive asymptotic decoding error probabilities are tolerated. For source coding, the strong converse establishes that irrespective of the allowed reconstruction error probabilities, a discrete-memoryless source cannot be compressed with a rate below the entropy of the source.  Similar  results were also established for generalized network scenarios \cite{fong2016proof,fong2017proof,tyagi2019strong}, i.e., for memoryless multi-user channels and  distributed compression systems \cite{GuEffros_1,GuEffros_2,oohama2018exponential,oohama2019exponential,kosut2018strong}. 

Our main interest in this paper is in distributed hypothesis testing problems where multiple terminals observe memoryless source sequences whose underlying joint distribution depends on a binary hypothesis $\mathcal{H}\in\{0,1\}$. Multiple decision centers wish to decide on the value of $\mathcal{H}$ based on their local source sequences and the communicated bits. Information-theorists showed great interest in Stein-setups where the type-I error probability (the probability of error under the null hypothesis $\mathcal{H}=0$) is required to stay asymptotically below a given threshold  in the infinite blocklength  regime,  while the type-II error probability (the probability of error under $\mathcal{H}=1$) has to decay to 0 exponentially fast with largest possible exponent \cite{Ahlswede,Han,shalaby,Amari,Wagner,Kim,Katz2,Gunduz_noisy_channels,Kochman,Gunduz_noisy_strong_converse_TAI,Michele_noisy_and_MAC,JSAIT,zhao2018distributed,Michele_MAC,ITW20,PierreMichele,HWS22_BC,salehkalaibar2020hypothesisv1,Vincent,GLOBECOM21,ITW21}. For  the two-terminal setup with  a single sensor communicating to a single decision center over a  rate-limited  link, Ahlswede and Csisz\'ar  \cite{Ahlswede} proved the strong converse result  that the largest possible type-II error exponent is independent of the admissible type-I error probability threshold. A similar strong converse result was also shown in the special case called ``testing against independence'' for  the two-hop hypothesis testing problem (see Figure~\ref{fig:Khop} for $K=2$) over rate-limited communication links where the last two terminals produce a guess on the binary hypothesis \cite{Vincent} and assuming  that the source sequences satisfy certain Markov chains. This latter strong converse result is based on a change of measure  and hyper-contractivity arguments \cite{liu2017_hyper_conc}. 

In this paper, we generalize the strong converse  result of \cite{Vincent} to an arbitrary number of $K$ hops. We thus show  that the set of  possible type-II error exponents  that are simultaneously achievable at the various decision centers when testing against independence in a $K$-hop system satisfying a given set of Markov chains, does not depend on the permissible type-I error probabilities and equals the region  under vanishing type-I error probabilities determined in \cite{salehkalaibar2020hypothesisv1}. The proof method applied in this paper relies on a similar change of measure argument as in \cite{tyagi2019strong, GuEffros_1,GuEffros_2}, where  we also restrict to jointly typical source sequences as \cite{GuEffros_1}. No variational characterizations,  blowing-up lemma\cite{MartonBU}, or hypercontractivity arguments are required. Instead, we rely on arguments showing that certain  Markov chains  hold in an asymptotic regime of infinite blocklengths. Our proof method seems to extend also to other applications. As  warm-ups we use our proof technique to establish the well-known strong converses  for  lossless and lossy compression  with side-information at the decoder.

\textit{Notation:}
We follow the notation in \cite{ElGamal} and use 
sans serif font for bit-strings: e.g., $\m$ for a deterministic and $\M$ for a random bit-string. We also use $\mathrm{len}(\m)$ to denote the length of a bit-string. 
Finally,   $\mathcal{T}_{\mu}^{(n)}(\cdot)$ denotes the strongly typical set as defined in \cite[Definition 2.8]{Csiszarbook}. 

		\section{Lossless Source Coding with Side-Information}

\subsection{Setup and Known Results}
Consider two terminals, an encoder observing the source sequence $X^n$ and a decoder observing the related side-information sequence $Y^n$, where we assume that 
	\begin{IEEEeqnarray}{rCl}
		(X^n,Y^n)  \textnormal{ i.i.d. } \sim \, P_{XY},
	\end{IEEEeqnarray} 
for a given probability mass function (pmf) $P_{XY}$ on the product alphabet $\mathcal{X}\times \mathcal{Y}$.  The encoder uses a function $\phi^{(n)}$ to compress the sequence $X^n$ into a bit-string message $\M$,
\begin{IEEEeqnarray}{rCl}
	\M&=& \phi^{(n)}(X^n) 
\end{IEEEeqnarray} 
of length $nR$ bits,  for a given rate $R>0$,
\begin{equation}
\len(\M)  = nR.
\end{equation}
Based on this message and its own observation $Y^n$, the decoder is supposed to reconstruct the source sequence $X^n$ with small probability of error. Thus, the decoder applies a decoding function $g^{(n)}$ to $(\M, Y^n)$ to produce the reconstruction sequence $\hat{X}^n \in \mathcal{X}^n$:
\begin{equation}
\hat{X}^n = g^{(n)}( \M, Y^n). 
\end{equation}

\begin{definition} Given $\epsilon \in [0,1)$. Rate $R>0$ is said $\epsilon$-achievable if there exist sequences (in $n$) of encoding and reconstruction functions $\phi^{(n)}$ and $g^{(n)}$  such that 
\begin{equation}\label{eq:error}
\varlimsup_{n\to \infty} \Pr[X^n \neq \hat{X}^n] \leq \epsilon.
\end{equation}
\end{definition}

A standard result in information theory says
\begin{theorem}
All  rates $R > H(X|Y)$ are $\epsilon$-achievable for all $\epsilon\in[0,1)$ and all rates $R<H(X|Y)$ are not $\epsilon$-achievable for any $\epsilon\in[0,1)$.
\end{theorem}
\medskip

In the following subsection we show  a new  converse proof. The goal is to illustrate (some of) the tools that we employ to prove our main result, Theorem~\ref{thm:fixedK} ahead.
\subsection{Alternative Strong Converse Proof}
Fix a sequence of encoding and decoding functions $\{\phi^{(n)}, g^{(n)}\}_{n=1}^\infty$ satisfying \eqref{eq:error}. 
We perform a similar  change of measure argument as in  \cite{tyagi2019strong,GuEffros_1} where we restrict to typical sequences. Define $\mu_n :=n^{-1/3}$ and the set 
\begin{equation} 
\mathcal{D}:= \left\{ (x^n, y^n) \in \mathcal{T}_{\mu_n}^{(n)}(P_{XY}) \colon g^{(n)}\left( \phi^{(n)}(x^n), y^n\right) =x^n\right\} , 
\end{equation}
i.e., the set of all typical $(x^n,y^n)$-sequences for which the reconstructed sequence $\hat{X}^n$ coincides with the source sequence $X^n$. 
Let $\Delta:=  \Pr[(X^n,Y^n) \in \mathcal{D}] $
and notice that by \eqref{eq:error} and \cite[Remark to Lemma~2.12]{Csiszarbook}:
\begin{equation} 
 \Delta\geq 1-\epsilon -\frac{|\mathcal{X}||\mathcal{Y}|}{4\mu_n^2 n},
\end{equation} 
and thus
\begin{equation} \label{eq:D}
\varliminf_{n\to \infty} \Delta\geq 1-\epsilon. 
\end{equation} 
Let  further $(\tilde{X}^n, \tilde{Y}^n)$  be random variables of joint pmf
\begin{equation} 
P_{\tilde{X}^n\tilde{Y}^n} (x^n,y^n) = \frac{P_{\tilde{X}^n\tilde{Y}^n} (x^n,y^n) }{\Delta} \cdot \mathds{1}\{ (x^n, y^n) \in \mathcal{D}\}.
\end{equation}
Let also $\tilde{\M}=\phi^{(n)}({\tilde{X}^n})$ and  $T$ be uniform over $\{1,\ldots, n\}$ independent of $(\tilde{X}^n, \tilde{Y}^n, \tilde{\M})$, and define $\tilde{X}:= \tilde{X}_T$ and  $\tilde{Y}:= \tilde{Y}_T$. 

Notice  the following sequence of equalities: 
\begin{IEEEeqnarray}{rCl}
\lefteqn{\frac{1}{n} H(\tilde{X}^n, \tilde{Y}^n) }  \nonumber \\ & = & - \frac{1}{n} \sum_{(x^n, y^n)\in \mathcal{D}} P_{\tilde{X}^n\tilde{Y}^n} (x^n,y^n) \log P_{\tilde{X}^n\tilde{Y}^n} (x^n,y^n) \label{eq:XY_step1}\\
& = & -\frac{1}{n}  \sum_{(x^n, y^n)\in \mathcal{D}} P_{\tilde{X}^n\tilde{Y}^n} (x^n,y^n) \log \frac{P_{{X}^n{Y}^n} (x^n,y^n)}{\Delta} \\
& = &  -  \frac{1}{n} \sum_{i=1}^n\sum_{(x^n, y^n)\in \mathcal{D}} P_{\tilde{X}^n\tilde{Y}^n} (x^n,y^n) \log P_{XY}(x_i,y_i) \nonumber\\
&& +\frac{1}{n} \log \Delta  \\
& = &  - \frac{1}{n} \sum_{i=1}^n \sum_{(x,y) \in \mathcal{X}\times \mathcal{Y}}  \hspace{-3mm}P_{\tilde{X}_i\tilde{Y}_i} (x,y) \log P_{XY}(x,y)+\frac{1}{n}\log \Delta\IEEEeqnarraynumspace \\
& = & -   \sum_{(x,y) \in \mathcal{X}\times \mathcal{Y}}  P_{\tilde{X}\tilde{Y}}(x,y) \log P_{XY}(x,y) + {1\over n} \log \Delta.  \IEEEeqnarraynumspace\label{eq:XY_last_step}
\end{IEEEeqnarray}
Since $(\tilde{X}^n, \tilde{Y}^n)\in \mathcal{T}_{\mu_n}^{(n)}(P_{XY})$, we have:
\begin{equation}
| P_{\tilde{X}\tilde{Y}}(x,y) - P_{{X}{Y}}(x,y)| \leq \mu_n,
\end{equation}
and thus as $n\to \infty$ (because  $\mu_n=n^{-1/3}$ and $\Delta$ is bounded away from $0$, see \eqref{eq:D}): 
\begin{equation} \label{eq:XY}
\lim_{n\to \infty} \frac{1}{n} H(\tilde{X}^n \tilde{Y}^n)  = H(XY). 
\end{equation}
In a similar manner one can obtain
\begin{equation} \label{eq:Y}
\lim_{n\to \infty} \frac{1}{n} H(\tilde{Y}^n)  = H(Y),
\end{equation} 
and thus combining \eqref{eq:XY} and \eqref{eq:Y}, by the chain rule:
\begin{equation} \label{eq:X_Y}
 \lim_{n\to \infty} \frac{1}{n} H(\tilde{X}^n|\tilde{Y}^n)  = H(X|Y). 
\end{equation}
The strong converse is then easily obtained by this Limit \eqref{eq:X_Y}, using the same steps as in the weak converse: 
\begin{IEEEeqnarray}{rCl}
R & \geq & \frac{1}{n} H(\tilde{\M}) \geq\frac{1}{n}    H(\tilde{\M}|\tilde{Y}^n)    \\
& = & \frac{1}{n}    I (\tilde{\M};\tilde{X}^n |\tilde{Y}^n)= \frac{1}{n}    H(\tilde{X}^n |\tilde{Y}^n), 
\end{IEEEeqnarray}
and  letting $n\to \infty$. Here, the last equality holds because by the definition of the set $\mathcal{D}$, the new source sequence $\tilde{X}^n$ can be obtained as a function of $\tilde{\M}$ and $\tilde{Y}^n$. 

\section{Lossy Source Coding with Side-Information}

\subsection{Setup and Known Results}
Reconsider the setup in the previous section, but with modified achievability definitions. Let a finite reconstruction alphabet ${\mathcal{Z}}$, a per-symbol distortion function $d\colon \mathcal{X} \times {\mathcal{Z}}\to \mathbb{R}_0^+$, and a maximum allowed distortion $D>0$ be given. 

\begin{definition} Given $\epsilon \in [0,1)$. Rate $R>0$ is said $\epsilon$-achievable if there exist sequences (in $n$) of encoding and reconstruction functions $\phi^{(n)}$ and $g^{(n)}$  such that  the excess distortion satisfies
\begin{equation}\label{eq:error2}
\varlimsup_{n\to \infty} \Pr\left[\frac{1}{n} \sum_{i=1}^n d(X_i\neq Z_i)> D \right ] \leq \epsilon.
\end{equation}
\end{definition}

From \cite{wynerziv} and \cite{oohama2018exponential} we have:
\begin{theorem}
All rates $R$ above  
\begin{equation}\label{eq:Rmin}
R_{\min}:= \min_{\substack{P_{S|X}, P_{Z|SY} \colon \\ E[d(X,Z)] \leq D } }I(X;S|Y),
\end{equation}
 are $\epsilon$-achievable for all $\epsilon\in[0,1)$ and all rates $R< R_{\min}$  are not $\epsilon$-achievable for any $\epsilon\in[0,1)$. Here, mutual information is calculated according to the pmf $P_{SXYZ}= P_{S|X}P_{XY}P_{Z|SY}$.
\end{theorem}

It is well known (and follows by convexity arguments) that in the minimization in \eqref{eq:Rmin} one can restrict to degenerate pmfs $P_{Z|SY}$ so that $Z$ is a function of $S$ and $Y$.

In the following section we present a new  proof of above strong converse, i.e., the non-achievability for any $R< R_{\min}$. 
\subsection{Alternative Strong Converse Proof}
Fix a sequence of encoding and decoding functions $\{\phi^{(n)}, g^{(n)}\}_{n=1}^\infty$ satisfying \eqref{eq:error}. 
We again perform a similar  change of measure argument as in  \cite{GuEffros_1}. Define $\mu_n :=n^{-1/3}$ and the set 
\begin{IEEEeqnarray}{rCl} 
\lefteqn{\mathcal{D}:= \bigg\{ (x^n, y^n) \in \mathcal{T}_{\mu_n}^{(n)}(P_{XY}) }\hspace{2cm}\nonumber \\
& \colon& d^{(n)}\left(x^n, g^{(n)}\left( \phi^{(n)}(x^n), y^n\right)\leq D\right)\bigg\} , \IEEEeqnarraynumspace
\end{IEEEeqnarray}
where we define the block-distortion function $d^{(n)}(x^n, \hat{x}^n):=\frac{1}{n}\sum_{i=1}^n d(\hat{x}_i,x_i)$. 
Let 
\begin{equation} \Delta:=  \Pr[(X^n,Y^n) \in \mathcal{D}] 
\end{equation}
and notice that by \eqref{eq:error2}  and \cite[Remark to Lemma~2.12]{Csiszarbook}
\begin{equation} \label{eq:D_WZ}
\varliminf_{n\to \infty} \Delta\geq 1-\epsilon. 
\end{equation} 
Let  further $(\tilde{X}^n, \tilde{Y}^n)$ be random variables of joint pmf
\begin{equation} 
P_{\tilde{X}^n\tilde{Y}^n} (x^n,y^n) = \frac{P_{\tilde{X}^n\tilde{Y}^n} (x^n,y^n) }{\Delta} \cdot \mathbbm{1}\{ (x^n, y^n) \in \mathcal{D}\}.
\end{equation}
Let also $\tilde{\M}=\phi^{(n)}\big(\tilde{X}^n\big)$,  $\tilde{Z}^n=g^{(n)}\big(\tilde{\M},\tilde{Y}^n\big)$, and  $T$ be uniform over $\{1,\ldots, n\}$ independent of $(\tilde{X}^n, \tilde{Y}^n, \tilde{\M},\tilde{Z}^n)$, and define $\tilde{X}:= \tilde{X}_T$,  $\tilde{Y}:= \tilde{Y}_T$, and $\tilde{Z}:= \tilde{Z}_T$

Following the steps in the weak converse, we have 
\begin{IEEEeqnarray}{rCl}
R & \geq & \frac{1}{n} H(\tilde{\M}) \geq\frac{1}{n}    H(\tilde{\M}|\tilde{Y}^n)    \\
& = & \frac{1}{n}    I (\tilde{\M};\tilde{X}^n |\tilde{Y}^n) =  \frac{1}{n}    \left[ H(\tilde{X}^n |\tilde{Y}^n) -  H(\tilde{X}^n |\tilde{Y}^n\tilde{\M}) \right]. \IEEEeqnarraynumspace
\end{IEEEeqnarray}
Let $\oi_1'(n)$ be an appropriate function tending to 0 as $n\to \infty$. Using similar arguments as leading to  \eqref{eq:X_Y}, where we use \eqref{eq:D_WZ} and the fact that $1-\epsilon>0$, we   continue to note
\begin{IEEEeqnarray}{rCl}
R & \geq & H(X|Y)+ \oi_1'(n) - \frac{1}{n}  \sum_{i=1}^{n} H(\tilde{X}_i |\tilde{X}^{i-1}\tilde{Y}_i \tilde{Y}_{i+1}^n \tilde{\M})\IEEEeqnarraynumspace  \\
& = & H(X|Y) + \oi_1'(n) -  H(\tilde{X}_T|\tilde{Y}_T \tilde U_{T}T)\\
& = & H(X|Y) + \oi_1'(n) -  H(\tilde{X}|\tilde{Y} U).\label{eq:M0}
\end{IEEEeqnarray}
where we set 
$\tilde U_i:=( \tilde{X}^{i-1},\tilde{Y}_{i+1}^n, \tilde{\M})$ and $U:=(\tilde U_T,T)$.

Since $\tilde{\M}$ is a function of $\tilde{X}^n$ and following similar steps as above (but where we exchange the roles of $\tilde{X}$ and $\tilde{Y}$), we obtain: 
\begin{IEEEeqnarray}{rCl} \label{eq:M1}
	0 & = &\frac{1}{n}  I(\tilde{\M};\tilde{Y}^n|\tilde{X}^n)
\\
& \geq & H(Y|X) + \oi_2'(n) -  H(\tilde{Y}_T|\tilde{X}_T \tilde U_{T}T) \\
& \geq &  H(Y|X) + \oi_2'(n) -  H(\tilde{Y}|\tilde{X}   U),\label{eq:M1last}
	\end{IEEEeqnarray}  
	where $ \oi_2'(n)$ is a function tending to 0 as $n\to \infty$.


The most complicated step is the following sequence of inequalities, where we use the telescoping identity in the same way as \cite{tyagi2019strong} in its strong converse proof for this problem. Since $\tilde{Z}^n$ is a function of $(\tilde{Y}^n,\tilde{\M})$, we have: 
\begin{IEEEeqnarray}{rCl} \label{eq:M2}
		0 & \geq &\frac{1}{n}    I ({\tilde{Z}^n};\tilde{X}^n |\tilde{Y}^n\tilde{\M}) \\
		& \stackrel{(a)}{=}  & \frac{1}{n}  \big[    H(\tilde{X}^n |\tilde{Y}^n\tilde{\M}) - H(\tilde{Y}^n |\tilde{X}^n\tilde{\M}) \nonumber \\
		&& \hspace{.5cm}+ H(\tilde{Y}^n |\tilde{X}^n) -  H(\tilde{X}^n |\tilde{Y}^n\tilde{\M}{\tilde{Z}}^n) \big] \\
				& \stackrel{(b)}{=} & \frac{1}{n}  \big[    H(\tilde{X}^n |\tilde{\M}) - H(\tilde{Y}^n |\tilde{\M}) \nonumber \\
				&& \hspace{.5cm}+ H(\tilde{Y}^n |\tilde{X}^n) -  H(\tilde{X}^n |\tilde{Y}^n\tilde{\M}{\tilde{Z}}^n) \big] \\
		&\stackrel{(c)}{\geq} & \frac{1}{n}    \sum_{i=1}^{n} \big[H(\tilde{X}_i |  \tilde{Y}_{i+1}^n,\tilde{X}^{i-1} \tilde{\M})- H(\tilde{Y}_i | \tilde{Y}_{i+1}^n,\tilde{X}^{i-1} \tilde{\M})\big] \nonumber \\
		&&+ \frac{1}{n} H(\tilde{Y}^n |\tilde{X}^n)- \frac{1}{n}   \sum_{i=1}^{n} H(\tilde{X}_i |\tilde{Y}_i, \tilde{Y}_{i+1}^n,\tilde{X}^{i-1} \tilde{\M}{\tilde{Z}}_i)\IEEEeqnarraynumspace \\
		& \stackrel{(d)}{= } & H(\tilde X|U) - H(\tilde Y|U) + H( Y|  X ) + \oi_2'(n)  - H(\tilde X|\tilde Y U \tilde{Z})  \nonumber \\\\
				& \stackrel{(e)}{=} & H(\tilde X|\tilde YU) - H(\tilde Y|\tilde XU) + H( Y|  X) + \oi_2'(n) \nonumber \\&& - H(\tilde X|\tilde Y U \tilde{Z}) \\ 
& = & I(\tilde X\tilde{Z}|\tilde YU) - H(\tilde Y|\tilde XU) + H( Y|  X) + \oi_2'(n) 
, \label{eq:fun}
		\end{IEEEeqnarray}  
		where $(a)$  holds because the mutual information $I(\tilde{Y}^n; \tilde{\M}|\tilde{X}^n)=0$; $(b)$ is obtained by adding and subtracting $I(\tilde{X}^n;\tilde{Y}^n |\tilde{\M})$; $(c)$ holds by a telescoping identity, by the chain rule, and because conditioning can only reduce entropy; $(d)$ holds by the definitions of the random variables $\tilde{X}, \tilde{Y}, \tilde{Z}$, and $U$ and by similar steps as leading to \eqref{eq:M1last}; and $(e)$ holds  by adding and subtracting $I(\tilde{X};\tilde{Y} |U)$. 
		
Finally, notice that by the definition of the set $\mathcal{D}$:
\begin{equation}\label{eq:D_WZ_2}
D \geq \E\left[ \frac{1}{n}\sum_{i=1}^n d(\tilde X_i,{\tilde Z}_i) \right]  = \E[ d(\tilde{X}, {\tilde{Z})}].
\end{equation}

 The desired  rate-upper bound $R\geq R_{\min}$  is then obtained by combining  \eqref{eq:M0}, \eqref{eq:M1last}, \eqref{eq:fun}, and \eqref{eq:D_WZ_2} and by taking $n\to \infty$. Details are as follows. By  Carath\'eodory's theorem \cite[Appendix C]{ElGamal}, there exists a set $\mathcal{U}$ of size 
\begin{align}
\vert {\mathcal{U}}\vert &\leq |\mathcal{X}|\cdot\vert \mathcal{Y}\vert \cdot\vert \mathcal{Z}\vert+ 1
\end{align}
so that for each $n$, we can find an auxiliary random variable ${U}$  taking value on $\mathcal{U}$ and so that \eqref{eq:M0}, \eqref{eq:M1last}, and \eqref{eq:fun} hold.
We  restrict to such auxiliary random variables  $U$  and invoke the Bolzano-Weierstrass theorem to conclude the existence of  a pmf $P_{U{X}{Y}{Z}}^*$ over $\mathcal{U}\times \mathcal{X}\times \mathcal{Y} \times \mathcal{Z}$, also abbreviated as $P^*$, and an increasing  subsequence of positive numbers $\{n_i\}_{i=1}^\infty$ so that 
	\begin{IEEEeqnarray}{rCl}\label{eq:ni}
		\lim_{{i\to \infty}} P_{U\tilde{X}\tilde{Y}\tilde{Z};n_i}&=& P_{U{X}{Y}{Z}}^*,
	\end{IEEEeqnarray}
	where $P_{U\tilde{X}\tilde{Y}\tilde{Z};n_i}$ denotes the pmf of the quadruple $(U,\tilde{X},\tilde{Y},\tilde{Z})$ at blocklength $n_i$. Notice that for any blocklength $n_i$ the pair $\big(\tilde{X}^{n_i},\tilde{Y}^{n_i}\big)$ lies in the jointly typical set $\mathcal{T}^{(n_i)}_{\mu_{n_i}}(P_{XY})$, i.e.,  {$\big\vert P_{\tilde X \tilde {Y}; n_i} - P_{XY}\big\vert \leq \mu_{n_i}$},  and thus by the definition of $(\tilde{X},\tilde{Y})$ and by \eqref{eq:ni}, the limiting pmf satisfies $P^*_{{X}{Y}}=P_{XY}$.

	By the monotone continuity  of mutual information over finite pmfs, we further deduce from \eqref{eq:M0}, \eqref{eq:M1last} and \eqref{eq:fun} that:
\begin{IEEEeqnarray}{rCl}		
	R& \geq & I_{P^*}(X;U|Y) \label{eq:Rr}\\
0 &= & I_{P^{*}}  I(Y;U|X) \\
0 &= & I_{P^{*}}  I(X;Z|YU),\label{eq:ffun}
	\end{IEEEeqnarray}
	where the subscript $P^*$ indicates that the mutual information quantities should be computed with respect to  $P^*$. More precisely, \eqref{eq:ffun} holds because as $i \to \infty$ the entropy $H(\tilde{Y}|\tilde{X}U)$ tends to $H_P^*({Y}|{X}U)$, and $H(Y|X)-H_P^*({Y}|{X}U) = I_{P^*}(U;Y|X)\geq 0$ because $P_{XY}^*=P_{XY}$.

Combined with \eqref{eq:D_WZ_2}, which implies:
\begin{IEEEeqnarray}{rCl}
D &\geq&   \E_{P^*}[ d({X}, {{Z}})],
\end{IEEEeqnarray}
above three (in)equalities \eqref{eq:Rr}--\eqref{eq:ffun}
prove that rates below $R_{\min}$ are not achievable for any $\epsilon\in[0,1)$.


	\section{Testing Against Independence in a $K$-Hop Network}
	Consider a system with a transmitter T$_0$ observing the source sequence $Y_0^n$, $K-1$ relays labelled $\text{R}_1,\ldots,\text{R}_{K-1} $ and observing sequences $Y_{1}^n, \ldots, Y_{K-1}^n$, respectively, and a receiver R$_{K}$ observing sequence $Y_{K}^n$.  

The source sequences $(Y_0^n,Y_1^n,\ldots,Y_{K}^n)$ are distributed according to one of two distributions depending on a binary hypothesis $\mathcal{H}\in\{0,1\}$:
\begin{subequations}\label{eq:dist_Khop}
	\begin{IEEEeqnarray}{rCl}
		& &\textnormal{if } \mathcal{H} = 0: (Y_0^n,Y_1^n,\ldots,Y_{K}^n)  \textnormal{ i.i.d. } \sim \, P_{Y_0Y_1\cdots Y_{K}}; \label{eq:H0_dist_Khop}\\
		& &\textnormal{if } \mathcal{H} = 1: (Y_0^n,Y_1^n,\ldots,Y_{K}^n)  \textnormal{ i.i.d. } \sim\, P_{Y_0}\cdot P_{Y_1}\cdots P_{Y_K}.\nonumber\\
	\end{IEEEeqnarray} 
\end{subequations}

\begin{figure}[htbp]
\vspace{-.7cm}
	\centerline{\includegraphics[ scale=0.52]{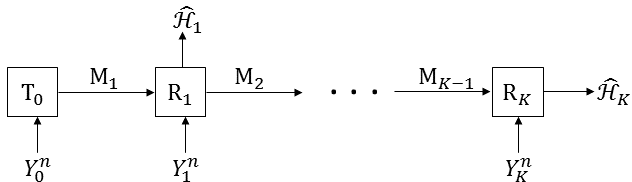}}
	\caption{Cascaded $K$-hop setup with $K$ decision centers.}
	\label{fig:Khop}
\end{figure}

Communication  takes place over $K$  hops as illustrated in Figure~\ref{fig:Khop}.  
The transmitter T$_{0}$  sends a message $\M_1 = \phi_0^{(n)}(Y_0^n)$ to the first relay R$_1$, which  sends a message $\M_2=\phi_1^{(n)}(Y_1^n,\M_1)$ to the second relay and so on. The communication is thus described by encoding functions 
\begin{IEEEeqnarray}{rCl}
	\phi_0^{(n)} &\colon& \mathcal{Y}_0^n \to \{0,1\}^\star \\
	\phi_k^{(n)} & \colon & \mathcal{Y}_k^n \times \{0,1\}^\star \to \{0,1\}^\star, \quad k\in\{1,\ldots, K-1\},\IEEEeqnarraynumspace 
\end{IEEEeqnarray} 
so that the produced message strings
\begin{IEEEeqnarray}{rCl}
	\M_1&=& \phi_0^{(n)}(\mathcal{Y}_0^n) \\
	\M_{k+1}& = & 	\phi_{k}^{(n)} ({Y}_k^n, \M_{k}) , \quad  k\in\{1,\ldots, K-1\},
\end{IEEEeqnarray} 
satisfy the maximum rate constraints 
\begin{equation}\label{eq:Ratek}
	\mathrm{len}\left(\M_{k}\right) \leq nR_{k}, \quad k \in \{1,\ldots,K\}.
\end{equation}

Each relay R$_1$, \ldots, R$_{K-1}$ as well as the receiver R$_K$, produces a guess of the hypothesis $\mathcal{H}$. 
These guesses are described by guessing functions 
\begin{equation}
	g_k^{(n)} \colon \mathcal{Y}_{k}^n \times \{0,1\}^{\star} \to \{0,1\}, \quad k\in\{1,\ldots, K\}, 
\end{equation}
where we request that the guesses
\begin{IEEEeqnarray}{rCl}
	\hat{\mathcal{H}}_{k,n} = g_k^{(n)}( Y_k^n, \M_{k}), 	\quad k\in\{1,\ldots, K\},
\end{IEEEeqnarray}  have type-I error probabilities 
\begin{IEEEeqnarray}{rCl}
	\alpha_{k,n} &\triangleq& \Pr[\hat{\mathcal{H}}_{k} = 1|\mathcal{H}=0], \quad k\in\{1,\ldots,K\},
\end{IEEEeqnarray}
not exceeding given thresholds $\epsilon_1,\epsilon_2,\ldots,\epsilon_K> 0$, and  type-II error probabilities
\begin{IEEEeqnarray}{rCl}
	\beta_{k,n} &\triangleq& \Pr[\hat{\mathcal{H}}_{k} = 0|\mathcal{H}=1],\quad  k\in\{1,\ldots,K\},
\end{IEEEeqnarray}
decaying to 0 exponentially fast with largest possible exponents.  

\medskip

\begin{definition} Given maximum type-I error probabilities $\epsilon_1,\epsilon_2,\ldots,\epsilon_K \in [0,1)$ and rates $R_1,R_2, \ldots, R_K \geq 0$. The exponent tuple $(\theta_1,\theta_2, \ldots, \theta_{K})$ is called \emph{$(\epsilon_1,\epsilon_2, \ldots, \epsilon_K)$-achievable} if there exists a sequence of encoding and decision functions $\big\{\phi_0^{(n)},\phi_1^{(n)}, \ldots, \phi_{K-1}^{(n)},g_1^{(n)},g_2^{(n)}, \ldots, g_K^{(n)}\big\}_{n\geq 1}$ satisfying for each $k \in \{1,\ldots,K\}$:
	\begin{subequations}\label{eq:Kconditions}
		\begin{IEEEeqnarray}{rCl}
	\text{len}(\M_k) &\leq& nR_k,\label{rate_constraint_Khop} \\
			\varlimsup_{n \to \infty}\alpha_{k,n} & \leq& \epsilon_k,\label{type1constraint1_Khop}\\ 
			\label{thetaconstraint_Khop}
			\varliminf_{n \to \infty}  {1 \over n} \log{1 \over \beta_{k,n}} &\geq& \theta_k.
		\end{IEEEeqnarray}
	\end{subequations}
\end{definition}
\begin{definition}\label{def:EregionK} 
	The fundamental exponents region $\mathcal{E}^*(R_1,R_2,\ldots,R_K,\epsilon_1,\epsilon_2,\ldots,\epsilon_K)$ is defined as the closure of the set of all $(\epsilon_1,\epsilon_2,\ldots,\epsilon_K)$-achievable exponent pairs $(\theta_{1},\theta_{2},\ldots,\theta_{K})$ for given  rates $R_1,\dots, R_K\geq 0$.
\end{definition}

\subsection{Previous Results on $K$-Hop Hypothesis Testing}\label{sec:maxresults_K}
	The  $K$-hop hypothesis testing setup of Figure~\ref{fig:Khop} and Equations~\eqref{eq:dist_Khop} was also considered in   \cite{salehkalaibar2020hypothesisv1} in the  special case   $\epsilon_1=\cdots =\epsilon_K=0$, for which the fundamental exponents region was determined. The result of \cite{salehkalaibar2020hypothesisv1}  is based on the following definition and presented in Theorem~\ref{thm:fixedK} ahead. 
	\begin{definition}
	For any $\ell\in\{1,\ldots, K\}$, define the function
	\begin{IEEEeqnarray}{rCl}
		\eta_\ell \colon  \mathbb{R}_0^+ & \to & \mathbb{R}_0^+ \\
		R & \mapsto&\max_{\substack{P_{U|Y_{\ell-1}}\colon \\R \geq I\left(U;Y_{\ell-1}\right)}} I\left(U;Y_\ell\right).
	\end{IEEEeqnarray}
\end{definition}
\medskip

	\begin{theorem}[Proposition 5 in \cite{salehkalaibar2020hypothesisv1}]\label{thm:fixedK_weak}	
		The fundamental exponents region 	satisfies
		\begin{IEEEeqnarray}{rCl}
		\lefteqn{\mathcal{E}^*(R_1,\ldots,R_K,0,\ldots,0) } \nonumber \\
		&=& \left\{(\theta_{1},\ldots, \theta_{K}) \colon \theta_k \leq \sum_{\ell=1}^{k}  \eta_\ell(R_\ell), \; k \in \{1,\ldots, K\}\right\} \IEEEeqnarraynumspace
		\end{IEEEeqnarray}
	\end{theorem}	

For $K=2$, Cao, Zhou, and Tan \cite{Vincent} also established the following strong converse result. 
	\begin{theorem}[Theorem 1 \cite{Vincent}]\label{thm:fixed2}	
		For $K=2$  and arbitrary $\epsilon_1, \epsilon_2\in[0,1)$:\vspace{-2mm}
		\begin{IEEEeqnarray}{rCl}
	{\mathcal{E}^*(R_1,R_2, \epsilon_1, \epsilon_2) } 
		&=& \left\{(\theta_{1}, \theta_{2}) \colon \theta_k \leq \sum_{\ell=1}^{k}  \eta_\ell(R_\ell), \, k \in \{1,2\}\right\} \nonumber\\
		\end{IEEEeqnarray}
	\end{theorem}	
\vspace{-3mm}

Our main result is a generalization of above Theorem~\ref{thm:fixed2} to an arbitrary number of  $K\geq 2$ hops. That is, we prove the strong converse to Theorem~\ref{thm:fixedK_weak}.

	\begin{theorem}\label{thm:fixedK}	
		For $K\geq 2$  and arbitrary $\epsilon_1, \ldots, \epsilon_K\geq 0$:
		\begin{IEEEeqnarray}{rCl}
\lefteqn{\mathcal{E}^*(R_1,\ldots,R_K,\epsilon_1,\ldots,\epsilon_K) } \; \nonumber \\
&= & \mathcal{E}^*(R_1,\ldots,R_K,0,\ldots,0) \\
	&=& \left\{(\theta_{1},\ldots, \theta_{K}) \colon \theta_k \leq \sum_{\ell=1}^{k}  \eta_\ell(R_\ell), \; k \in \{1,\ldots, K\}\right\} \IEEEeqnarraynumspace \label{eq:res_thetak}
		\end{IEEEeqnarray}
	\end{theorem}	
	\begin{IEEEproof}
	See the following Section~\ref{sec:proof}.
	\end{IEEEproof}
	
	\section{Strong Converse Proof of Theorem~\ref{thm:fixedK}}\label{sec:proof}
	Fix an exponent-tuple $(\theta_1,\ldots, \theta_K)$ in the exponents region $\mathcal{E}^*(R_1,\ldots, R_K, \epsilon_1,\ldots, \epsilon_K)$, and a sequence (in $n$) of encoding and decision functions $\{(\phi_0^{(n)}, \phi_1^{(n)}, \ldots, \phi_{K-1}^{(n)}, g_1^{(n)}, \ldots, g_K^{(n)})\}_{n\geq 1}$ achieving this tuple, i.e., satisfying constraints \eqref{eq:Kconditions}. 

Fix an arbitrary $k\in\{1,\ldots, K\}$  and set ${\mu_n=n^{-1/3}}$. Let $\mathcal{A}_{k}$ denote the acceptance region of R$_k$, i.e., 
\begin{equation}\label{eq:acceptance_region_k}
\mathcal{A}_k:= \{ (y_0^n, \ldots, y_{k}^n) \colon g_k^{(n)}(\m_k,y_k^n)=0\},
\end{equation}
where we define recursively $\m_1:=\phi_0^{(n)}(y_0^n)$ and 
\begin{equation}
\m_\ell := \phi_{\ell-1}^{(n)}(\m_{\ell-1},y_{\ell-1}), \ell\in\{ 2, \ldots, k\}.
\end{equation}
Define also  the intersection of this acceptance region with the typical set:  
\begin{IEEEeqnarray}{rCl}\label{B_Khop_strong_conv}
	\mathcal{D}_{k}\triangleq  \mathcal{A}_k \cap \mathcal{T}_{\mu_n}^{(n)}(P_{Y_0\cdots Y_{k}}).
\end{IEEEeqnarray}
By \cite[Remark to Lemma~2.12]{Csiszarbook} and the type-I error probability constraints in \eqref{type1constraint1_Khop}, 
\begin{IEEEeqnarray}{rCl}\label{eq:PB_KHop_strong_conv}
	\Delta_k:= P_{Y_0^nY_1^n\cdots Y_{k}^n}(\mathcal{D}_{k}) &\geq& 1 - \epsilon_k - {\vert{\mathcal{Y}_0}\vert \cdots\vert{\mathcal{Y}_{k}}\vert \over{{4 \mu_n^2} n}}, 
\end{IEEEeqnarray}
and thus  $\varliminf_{n\to \infty} \Delta_k \geq  1-\epsilon_k>0$ as $n\to \infty$. 

Let   $(\tilde{Y}_0^n, \tilde{Y}_1^n, \ldots, \tilde{Y}_k^n)$ be random variables of joint pmf
\begin{IEEEeqnarray}{rCl}
\lefteqn{P_{\tilde{Y}_0^n, \tilde{Y}_1^n, \ldots, \tilde{Y}_k^n} (\tilde{y}_0^n, \tilde{y}_1^n, \ldots, \tilde{y}_k^n) } \nonumber \\
& = & \frac{P_{\tilde{Y}_0^n, \tilde{Y}_1^n, \ldots, \tilde{Y}_k^n}(\tilde{y}_0^n, \tilde{y}_1^n, \ldots, \tilde{y}_k^n) }{\Delta} \cdot \mathds{1}\{ (\tilde{y}_0^n, \tilde{y}_1^n, \ldots, \tilde{y}_k^n) \in \mathcal{D}_k\}. \nonumber \\
\end{IEEEeqnarray}
Let also $\tilde{\M}_\ell=\phi_{\ell-1}^{(n)}(\tilde{\M}_{\ell-1},\tilde{Y}_{\ell-1}^n)$ and  $T$ be uniform over $\{1,\ldots, n\}$ independent of $(\tilde{Y}_0^n, \tilde{Y}_1^n, \ldots, \tilde{Y}_k^n, \tilde{\M}_1,\ldots, \tilde{\M}_k)$, and define $\tilde{Y}_\ell:= \tilde{Y}_{\ell,T}$ for $\ell\in\{1,\ldots, k\}$.

\medskip

At the end of this section, we prove the following Lemma~\ref{lem:strong}.
\begin{lemma}\label{lem:strong}
There exist random variables $\{U_{1}, \ldots, U_{k}\}$  satisfying  the (in)equalities 
\begin{subequations}\label{eq:conditions_K_strong_converse}
	\begin{IEEEeqnarray}{rCl}
		nR_\ell \geq H(\tilde{\M}_{\ell}) &\geq& nI(U_{\ell};\tilde{Y}_{\ell-1}) + \log \Delta_{k},  \quad \ell \in\{1,\ldots, k\}, \label{eq:Mji_Khop_strong_conv} \nonumber \\ \\
		I(U_{\ell}; \tilde{Y}_{\ell}|\tilde{Y}_{\ell-1}) & = & \oi_{1,\ell}(n), \label{eq:MarkovK_strong_conv}
	\end{IEEEeqnarray}
	and 
	\begin{IEEEeqnarray}{rCl}
		\lefteqn{-{1\over n}\log\Pr[\hat{\mathcal{H}}_{k} = 0| \mathcal{H}=1,(Y_0^n,\ldots,Y_{k}^n)\in\D_{k}] } \nonumber \\
		& \leq& \sum_{\substack{\ell =1}}^k I(U_{\ell};\tilde{Y}_{\ell})+ \oi_{2}(n), \hspace{4.5cm} \label{eq:theta_ub_k}
	\end{IEEEeqnarray}
\end{subequations}
where    the functions $\{\oi_{1,\ell}(n)\}_{\ell=1}^k$ and $\oi_{2}(n)$ all tend to 0 as $n\to \infty$.
\end{lemma} 

\medskip
 The desired bound on $\theta_k$ in \eqref{eq:res_thetak} is then obtained from above lemma by taking $n\to \infty$, as we explain in the following. 	
By  Carath\'eodory's theorem \cite[Appendix C]{ElGamal}, for each $n$ there must exist random variables ${U}_{1},\ldots ,U_{k}$ satisfying \eqref{eq:conditions_K_strong_converse} over alphabets of sizes
\begin{align}
\vert {\mathcal{U}}_{\ell} \vert &\leq |\mathcal{Y}_{\ell-1}|\cdot\vert \mathcal{Y}_\ell\vert + 2, \qquad \ell \in\{1,\ldots, k\}.
\end{align}
We thus restrict to random variables of above (bounded) supports and invoke the Bolzano-Weierstrass theorem to conclude the existence of  a pmf $P^{(\ell)}_{Y_{\ell-1}Y_{\ell}{U_{\ell}}}$ over $\mathcal{Y}_{\ell-1}\times \mathcal{Y}_{\ell} \times {\mathcal{U}_{\ell}}$, also abbreviated as $P^{(\ell)}$, and an increasing  subsequence of positive numbers $\{n_i\}_{i=1}^\infty$ satisfying
	\begin{IEEEeqnarray}{rCl}
		\lim_{i\to \infty} P_{\tilde{Y}_{\ell-1}\tilde{Y}_{\ell}{U}_{\ell};n_i}&=& P^{(\ell)}_{Y_{\ell-1}Y_{\ell}{U_{\ell}}}, \quad \ell \in\{1,\ldots, k\},\IEEEeqnarraynumspace
	\end{IEEEeqnarray}
	where $P_{\tilde{Y}_{\ell-1}\tilde{Y}_{\ell}{U}_{\ell};n_i}$ denotes the pmf at blocklength $n_i$. 
	
	By the monotone continuity  of mutual information over finite pmfs, we can then deduce that
\begin{IEEEeqnarray}{rCl}		\label{eq:R1}	
	R_\ell& \geq & I_{P^{(\ell)}}({U}_{\ell};{Y}_{\ell-1}), \quad \ell \in\{1,\ldots, k\},\\
	\theta_k &\leq &\sum_{\ell=1}^k I_{P^{(\ell)}}({U}_{\ell};{Y}_{\ell}), \label{theta_2_f}
	\end{IEEEeqnarray}
	where the subscripts indicate that mutual informations should be computed according to the indicated pmfs. 

Since for any blocklength $n_i$ the pair $\big(\tilde{Y}_{\ell-1}^{n_i},\tilde{Y}_{\ell}^{n_i}\big)$ lies in the jointly typical set $\mathcal{T}^{(n_i)}_{\mu_{n_i}}(P_{Y_{\ell-1}Y_\ell})$, we have  {$\big\vert P_{{Y}_{\ell-1}{Y}_{\ell}; n_i} - P_{{Y_{\ell-1}}Y_{\ell}}\big\vert \leq \mu_{n_k}$} and thus the limiting pmfs satisfy $P^{(\ell)}_{Y_{\ell-1}Y_{\ell}}=P_{Y_{\ell-1}Y_{\ell}}$. 
By similar continuity considerations and by \eqref{eq:MarkovK_strong_conv}, for all $\ell\in\{1,\ldots, k\}$   the Markov chain
\begin{IEEEeqnarray}{rCl}\label{eq:MC2_1_genconv}
	U_{\ell}\to Y_{\ell-1} \to Y_{\ell}, 
\end{IEEEeqnarray}	
holds  under $P_{Y_{\ell-1}Y_{\ell}U_{\ell}}^{(\ell)}$.

By the definitions of the functions $\{\eta_\ell(\cdot)\}$ and by \eqref{eq:R1}--\eqref{eq:MC2_1_genconv}:
	\begin{IEEEeqnarray}{rCl}\label{eq:E3_generalconverse}
		\theta_{k} &\leq&  \sum_{\ell=1}^k \eta_\ell(R_{\ell}),
	\end{IEEEeqnarray}
	which concludes the proof.

\medskip
\subsection{Proof of Lemma~\ref{lem:strong}}

Define  $\tilde{U}_{\ell,t}\triangleq(\tilde{\M}_\ell,\tilde{Y}_0^{t-1},\ldots,\tilde{Y}_{k}^{t-1})$ for $\ell \in \{1,\ldots,k\}$ and notice:
\begin{IEEEeqnarray}{rCl}
	H(\tilde{\M}_\ell)	&=& I(\tilde{\M}_\ell;\tilde{Y}_0^n\cdots \tilde{Y}_{k}^n)  
	\IEEEeqnarraynumspace\label{m1entropylbstep1_lemmaKHop}\\
	&=& H(\tilde{Y}_0^n\cdots\tilde{Y}_{k}^n) 
	 -  H(\tilde{Y}_0^n\cdots \tilde{Y}_{k}^n|\tilde{\M}_{\ell}) \IEEEeqnarraynumspace\\
	&=& n H(\tilde{Y}_{0,T}\cdots\tilde{Y}_{k,T}) + \log\Delta_{k} + \oi_1(n) \nonumber \\
	 && - \sum_{t=1}^{n} H(\tilde{Y}_{0,t}\cdots \tilde{Y}_{k,t}|\tilde{U}_{\ell,t})\IEEEeqnarraynumspace\label{m1entropylbstep4_lemmaKHop}\\
	&=& n H(\tilde{Y}_{0,T}\cdots \tilde{Y}_{k,T}) \log\Delta_{k} + \oi_1(n) \nonumber \\
	&& \quad - n H(\tilde{Y}_{0,T}\cdots \tilde{Y}_{k,T}|\tilde{U}_{\ell,T},T)] \label{Tuniformdef_lemmaKHop}\\
	&=& n [I(\tilde{Y}_0\cdots\tilde{Y}_{k};U_{\ell})] +  \log{\Delta_{k}} + \oi_1(n)  \label{eq:HM1_LB_lemma_last_eqKHop} \\
	&\geq& n \left[I(\tilde{Y}_{\ell-1};U_{\ell}) + {1 \over n} \log{\Delta_{k}}\right] + \oi_1(n).\IEEEeqnarraynumspace \label{eq:HM1_LB_lemmaKHop}
\end{IEEEeqnarray}
Here, (\ref{m1entropylbstep4_lemmaKHop}) holds by similar steps to \eqref{eq:XY_step1}--\eqref{eq:XY_last_step}, where $\oi_1(n)$ 
 is a function that tends to $0$ as $ n \to \infty$, by the chain rule, by the definition of $\tilde{U}_{\ell,t}$, and  by defining $T$ uniform over $\{1,\dots,n\}$ independent of all other random variables; and \eqref{eq:HM1_LB_lemma_last_eqKHop} holds by defining $U_\ell \triangleq (\tilde{U}_{\ell,T},T)$ and $\tilde{Y}_\ell\triangleq \tilde{Y}_{\ell,T}$ for all $\ell \in \{0,\ldots,k\}$.\\
This proves Inequality \eqref{eq:Mji_Khop_strong_conv} in the lemma.

We next upper bound the type-II error exponent $\theta_{k}$.
Define:
\begin{IEEEeqnarray}{rCl}
Q_{\tilde{\M}_k}(\m_k)&\triangleq& \sum_{y_0^n,y_1^n,\ldots, y_{k-1}^n}   P_{\tilde{Y}_{0}^n}(y_0^n)\cdots P_{\tilde{Y}_{k-1}^n}(y_{k-1}^n)  \nonumber\\
	&&  \quad \cdot \mathds{1}\{\m_k=\phi_k(\phi_{k-1}(\cdots(\phi_1(y_0^n)\cdots)), y_{k-1}^n)\}, \nonumber\\
\end{IEEEeqnarray}
and 
\begin{IEEEeqnarray}{rCl}
Q_{{\M}_k}(\m_K) 
	&\triangleq& \sum_{y_0^n,y_1^n,\ldots, y_{k-1}^n}   P_{{Y}_{0}^n}(y_0^n)\cdots P_{{Y}_{k-1}^n}(y_{k-1}^n)  \nonumber\\
	&&  \;\; \cdot \mathds{1}\{\m_k=\phi_{k-1}(\phi_{k-2}(\cdots(\phi_0(y_0^n)\cdots)), y_{k-1}^n)\}. \nonumber\\
\end{IEEEeqnarray}
and notice that
\begin{IEEEeqnarray}{rCl}
	Q_{\tilde{\M}_k}P_{\tilde{Y}_k^n}\left({\mathcal{A}}_{k}\right)
	&\leq& Q_{{\M}_k}P_{{Y}_k^n}\left({\mathcal{A}}_{k}\right)\Delta_k^{-(k+1)}
= \beta_{k,n} \Delta_k^{-(k+1)}\IEEEeqnarraynumspace \label{eq:11KHop}
\end{IEEEeqnarray}
Notice that by  \eqref{eq:acceptance_region_k}, the probability $P_{\tilde{\M}_k\tilde{Y}_k^n}({\mathcal{A}}_{k})=1$, and thus by 
 \eqref{eq:11KHop}  and standard inequalities (see \cite[Lemma~1]{JSAIT}):
\begin{IEEEeqnarray}{rCl}
	-{1\over n}\log \beta_{k,n} & \leq &-{1\over n}\log \left( Q_{\tilde{\M}_k}P_{\tilde{Y}_k^n}\left({\mathcal{A}}_{k}\right)  \right) - \frac{(k+1)}{n} \log \Delta_k \nonumber\\\\
	&\leq& {1 \over n } D(P_{\tilde{\M}_k\tilde{Y}_k^n}||Q_{\tilde{\M}_k}P_{\tilde{Y}_k^n}) + \delta_n'\IEEEeqnarraynumspace\label{theta_ub_lemma_relayKHop}
\end{IEEEeqnarray}
where $\delta_n'\triangleq - \frac{(k+1)}{n} \log \Delta_k +\frac{1}{n}$ and tends to 0 as $n \to \infty$.

We continue to upper bound the divergence term as
\begin{IEEEeqnarray}{rCl}
	\lefteqn{D(P_{\tilde{\M}_k\tilde{Y}_k^n}||Q_{\tilde{\M}_k}P_{\tilde{Y}_k^n})}\qquad \nonumber\\
	&=& I(\tilde{\M}_k;\tilde{Y}_k^n) + D(P_{\tilde{\M}_k}||Q_{\tilde{\M}_k}) \\
	&\leq& I(\tilde{\M}_k;\tilde{Y}_k^n) + D(P_{\tilde{Y}_{k-1}^n\tilde{\M}_{k-1}}||P_{\tilde{Y}_{k-1}^n}Q_{\tilde{\M}_{k-1}})\label{eq:dp_ineq_relative_entropyKHop}\\
	&\leq& I(\tilde{\M}_k;\tilde{Y}_{k}^n) + I(\tilde{\M}_{k-1};\tilde{Y}_{k-1}^n) \nonumber \\
	&& \qquad \qquad \qquad + D(P_{\tilde{Y}_{k-2}^n\tilde{\M}_{k-2}}||P_{\tilde{Y}_{k-2}^n}Q_{\tilde{\M}_{k-2}})\IEEEeqnarraynumspace\\
	&\vdots& \nonumber\\
	&\leq& \sum_{\ell=1}^{k} I(\tilde{\M}_\ell;\tilde{Y}_{\ell}^n)\\
	&\leq& \sum_{\ell=1}^k \sum_{t=1}^n I(\tilde{\M}_\ell\tilde{Y}_0^{t-1}\cdots \tilde{Y}_{k}^{t-1};\tilde{Y}_{\ell,t})\IEEEeqnarraynumspace\label{eq:divergence_markovcahinsKHop}\\
	&=& \sum_{\ell=1}^k \sum_{t=1}^n I(\tilde{U}_{\ell,t};\tilde{Y}_{\ell,t})\label{eq:divergence_end1KHop}\\
	&\leq&  n \sum_{\ell=1}^k I(U_\ell;\tilde{Y}_\ell) \label{theta_ub2_lemmaKHop}.
\end{IEEEeqnarray}
Here \eqref{eq:dp_ineq_relative_entropyKHop} is obtained by the data processing inequality for KL-divergence 
and \eqref{eq:divergence_end1KHop}--\eqref{theta_ub2_lemmaKHop} by the definitions of $\tilde{U}_{\ell,t},U_\ell,\tilde{Y}_\ell$ and $T$.

Combined with \eqref{theta_ub_lemma_relayKHop} this establishes Inequality~\eqref{eq:theta_ub_k}.

Finally, we proceed to prove that for any $\ell\in\{1,\ldots, k\}$  the Markov chain $U_\ell \to \tilde{Y}_{\ell-1} \to \tilde{Y}_{\ell}$ holds in the limit as $n \to \infty$. We start by noticing the Markov chain $\tilde{\M}_1 \to \tilde{Y}_0^n \to (\tilde{Y}_1^n,\cdots,\tilde{Y}_k^n)$, and thus:
\begin{IEEEeqnarray}{rCl}
	0 &=& I(\tilde{\M}_1;\tilde{Y}_1^n\cdots \tilde{Y}_{k}^n|\tilde{Y}_0^n) \label{MC1proofstep0KHop}\\ 
	&=& H(\tilde{Y}_1^n\cdots \tilde{Y}_{k}^n|\tilde{Y}_0^n)  - H(\tilde{Y}_1^n\cdots \tilde{Y}_{k}^n|\tilde{Y}_0^n\tilde{\M}_1)
	\label{MC1proofstep1KHop}\\
	&=& nH(\tilde{Y}_{1,T}\cdots \tilde{Y}_{k,T}|\tilde{Y}_{0,T}) 
	+\log{\Delta_{k}} +\tilde{\oi}_1(n) \nonumber\\
	&& - H(\tilde{Y}_1^n \cdots \tilde{Y}_{k}^n|\tilde{Y}_0^n\tilde{\M}_1) \label{MC1proofstep2KHop}\\
	&\geq & nH(\tilde{Y}_{1,T}\cdots\tilde{Y}_{k,T}|\tilde{Y}_{0,T}) 
	+\log{\Delta_{k}}+\tilde{\oi}_1(n) \nonumber \\
	&&-  n H(\tilde{Y}_{1,T}\cdots \tilde{Y}_{k,T} |\tilde{Y}_{0,T}\tilde{Y}_0^{T-1}\cdots \tilde{Y}_{k}^{T-1}\tilde{Y}_{0,T+1}^n\tilde{\M}_1T)\label{MC1proofstep4KHop}\IEEEeqnarraynumspace\\
	&=& nI(\tilde{Y}_{1,T}\cdots \tilde{Y}_{k,T};\tilde{Y}_0^{T-1}\cdots \tilde{Y}_{k}^{T-1}\tilde{Y}_{0,T+1}^n\tilde{\M}_1T|\tilde{Y}_{0,T}) \nonumber \\
	&&   + \log{\Delta_{k}} +\tilde{\oi}_1(n) \\
	&\geq& nI(\tilde{Y}_{1}\cdots \tilde{Y}_{k};{U}_1|\tilde{Y}_0)  + \log{\Delta_{k}}+  \tilde{\oi}_1(n),\label{MC1proofstep5KHop}
\end{IEEEeqnarray}
for some function $\tilde{\oi}_1(n)$ so that $\frac{1}{n} \tilde{\oi}_1(n)$ tends to 0 as $n \to \infty$, and where \eqref{MC1proofstep2KHop} can be shown in a similar manner to \eqref{eq:X_Y} 
and \eqref{MC1proofstep5KHop} by the definitions of $\tilde{Y}_\ell$, $\tilde{Y}_0$, $\tilde{U}_{1,t}$, and $U_1$ for all $\ell \in \{1,\ldots,k\}$. 

Since $\Delta_k$ is bounded,  $\frac{1}{n}\log \Delta_k$ tends to 0 as $n\to \infty$, and we can conclude that
\begin{equation}
\lim_{n\to \infty}I(\tilde{Y}_{1}\cdots \tilde{Y}_{k};\tilde{U}_1|\tilde{Y}_0)=0,
\end{equation}
thus proving \eqref{eq:MarkovK_strong_conv} for $\ell=1$. 

Notice next that for any $\ell\in\{2,\ldots, k\}$:
\begin{IEEEeqnarray}{rCl}
	I(U_\ell;\tilde{Y}_{\ell}| \tilde{Y}_{\ell-1}) &\leq& I(U_\ell\tilde{Y}_0\cdots \tilde{Y}_{\ell-2};\tilde{Y}_{\ell}| \tilde{Y}_{\ell-1})\\
	&=& I(U_\ell;\tilde{Y}_{\ell}| \tilde{Y}_0 \cdots \tilde{Y}_{\ell-1}) \nonumber \\
	&& + I(\tilde{Y}_0\cdots \tilde{Y}_{\ell-2};\tilde{Y}_{\ell}| \tilde{Y}_{\ell-1}) . \label{eq:sum}\IEEEeqnarraynumspace
\end{IEEEeqnarray}
In the following we show that both quantities $I(U_\ell;\tilde{Y}_{\ell}| \tilde{Y}_0 \cdots \tilde{Y}_{\ell-1})$ and $I(\tilde{Y}_0\cdots \tilde{Y}_{\ell-2};\tilde{Y}_{\ell}| \tilde{Y}_{\ell-1})$ tend to 0 as $n \to \infty$, which  establishes  \eqref{eq:MarkovK_strong_conv} for $\ell\in\{2,\ldots, k\}$. 

To prove that $I(\tilde{Y}_0\cdots \tilde{Y}_{\ell-2};\tilde{Y}_{\ell}| \tilde{Y}_{\ell-1})$ tends to 0 as $n \to \infty$, we notice that for any $\ell\in\{2,\ldots, k\}$:
\begin{IEEEeqnarray}{rCl}
D(P_{\tilde{Y}_{0}\cdots \tilde{Y}_{k}}||P_{Y_0\cdots Y_{k}})&\geq&  D(P_{\tilde{Y}_{0}\cdots \tilde{Y}_{\ell}}||P_{Y_0\cdots Y_{\ell}}) \\
	&=& D(P_{\tilde{Y}_{0}\cdots \tilde{Y}_{\ell}}||P_{Y_0\cdots Y_{\ell-1}}P_{Y_{\ell}|Y_{\ell-1}}) \\
	&= &  D(P_{\tilde{Y}_{0}\cdots \tilde{Y}_{\ell-1}}|| P_{\tilde{Y}_0\cdots\tilde{Y}_{\ell-1} }P_{\tilde{Y}_{\ell}|\tilde{Y}_{\ell-1}} ) \nonumber \\ 
	& & +\mathbb{E}_{P_{\tilde{Y}_{\ell-1}} }\left[ D(P_{\tilde{Y}_{\ell}|\tilde{Y}_{\ell-1}}||P_{{Y}_{\ell}| {Y}_{\ell-1}}) \right] \nonumber \\
	&& \qquad + D(P_{\tilde{Y}_0\cdots\tilde{Y}_{\ell-1}} ||P_{Y_0\cdots Y_{\ell-1}})    \\
	&\geq & D(P_{\tilde{Y}_{0}\cdots \tilde{Y}_{\ell}}|| P_{\tilde{Y}_0\cdots\tilde{Y}_{\ell-1}} P_{\tilde{Y}_{\ell}|\tilde{Y}_{\ell-1}} ) \IEEEeqnarraynumspace\\
	&\geq& I(\tilde{Y}_0\cdots \tilde{Y}_{\ell-2};\tilde{Y}_{\ell}| \tilde{Y}_{\ell-1}) \label{eq:la}.
\end{IEEEeqnarray}
Since $(\tilde{Y}_0\cdots \tilde{Y}_{k})$ lie in the jointly typical set $\mathcal{T}_{\mu_n}^{(n)}(P_{Y_0\cdots Y_{k}})$:
\begin{equation}|P_{\tilde{Y}_0\cdots \tilde{Y}_{k}} - P_{Y_0\cdots Y_{k}}| \leq \mu_n
	.
\end{equation}
Recalling that $\mu_n \downarrow 0$ as $n \to \infty$,  and by the continuity of the KL-divergence, we conclude that  $D(P_{\tilde{Y}_{0}\cdots \tilde{Y}_{k}}||P_{Y_0\cdots Y_{k}})$ tends to 0 as $n\to \infty$, and thus by \eqref{eq:la} and the nonnegativity of mutual information:
\begin{IEEEeqnarray}{rCl}\label{eq:summand1}
	\lim_{n\to \infty} I(\tilde{Y}_0\cdots \tilde{Y}_{\ell-2};\tilde{Y}_{\ell}| \tilde{Y}_{\ell-1}) =0.
\end{IEEEeqnarray}


Following similar steps to \eqref{MC1proofstep0KHop}--\eqref{MC1proofstep5KHop}, we further obtain: 
\begin{IEEEeqnarray}{rCl}
	0 &=& I(\tilde{\M}_\ell;\tilde{Y}_\ell^n\cdots \tilde{Y}_{k}^n|\tilde{Y}_0^n\cdots\tilde{Y}_{\ell-1}^n ) \nonumber \\
	& = & H(\tilde{Y}_\ell^n\cdots \tilde{Y}_{K}^n|\tilde{Y}_0^n\cdots\tilde{Y}_{k-1}^n ) \nonumber\\
	&&  -H(\tilde{Y}_k^n\cdots \tilde{Y}_{k}^n|\tilde{Y}_0^n\cdots\tilde{Y}_{\ell-1}^n \tilde{\M}_\ell)\\
%
	&=& nH(\tilde{Y}_{\ell,T}\cdots \tilde{Y}_{k,T}|\tilde{Y}_{0,T}\cdots \tilde{Y}_{\ell-1,T})
	+\log{\Delta_{k}} + \tilde{\oi}_{\ell}(n) \nonumber\\
	&& - H(\tilde{Y}_\ell^n\cdots \tilde{Y}_{k}^n|\tilde{Y}_0^n\cdots \tilde{Y}_{\ell-1}^n \tilde{\M}_\ell) \label{MC1proofstep2KHop_MC2}\\
	&\geq& n H(\tilde{Y}_{\ell,T} \cdots \tilde{Y}_{k,T} |\tilde{Y}_{0,T} \cdots \tilde{Y}_{\ell-1,T}) 
	+\log{\Delta_{k}} + \tilde{\oi}_{\ell}(n) \nonumber \\ 	
	&&-  \sum_{t=1}^{n}H(\tilde{Y}_{\ell,t} \cdots \tilde{Y}_{k,t} |\tilde{Y}_{0,t}\cdots \tilde{Y}_{\ell-1,t} \nonumber \\
	&& \qquad \qquad \tilde{Y}_0^{t-1}\cdots \tilde{Y}_{k}^{t-1} \tilde{Y}_{0,t+1}^n \cdots \tilde{Y}_{\ell-1,t+1}^n \tilde{\M}_\ell)\label{MC1proofstep3KHop_MC2}\IEEEeqnarraynumspace\\
	&=&  n H(\tilde{Y}_{\ell,T} \cdots \tilde{Y}_{k,T} |\tilde{Y}_{0,T} \cdots \tilde{Y}_{\ell-1,T}) +\log{\Delta_{k}} + \tilde{\oi}_{\ell}(n) \nonumber \\ 
	&&- n H(\tilde{Y}_{\ell,T} \cdots \tilde{Y}_{k,T} |\tilde{Y}_{0,T}\cdots \tilde{Y}_{\ell-1,T} \nonumber \\
	&& \qquad \qquad \tilde{Y}_0^{T-1}\cdots \tilde{Y}_{k}^{T-1} \tilde{Y}_{0,T+1}^n \cdots \tilde{Y}_{\ell-1,T+1}^n \tilde{\M}_\ell T)\label{MC1proofstep5KHop_MC2}\\
	&\geq& nI(\tilde{Y}_{\ell,T}\cdots \tilde{Y}_{k,T};\tilde{Y}_0^{T-1}\cdots \tilde{Y}_{k}^{T-1}\tilde{\M}_{\ell}T|\tilde{Y}_{0,T}\cdots \tilde{Y}_{\ell-1,T} ) \nonumber \\
	&& \qquad \quad + \log{\Delta_{k}} + \tilde{\oi}_{\ell}(n)\\
	&=& nI(\tilde{Y}_{\ell}\cdots \tilde{Y}_{k};U_\ell|\tilde{Y}_{0}\cdots \tilde{Y}_{\ell-1} ) + \log{\Delta_{k}} + \tilde{\oi}_{\ell}(n),
	\IEEEeqnarraynumspace\label{MC1proofstep6KHop_MC2}
\end{IEEEeqnarray}
where $\tilde{\oi}_\ell(n)$ is a function so that $\frac{1}{n} \tilde{\oi}_\ell(n)$  tends to 0 as $n\to \infty$.
Since $\Delta_k$ is bounded,  $\frac{1}{n}\log \Delta_k$ tends to 0 as $n\to \infty$, we can conclude that
\begin{equation}
\lim_{n\to \infty}I(\tilde{Y}_{\ell};\tilde{U}_\ell|\tilde{Y}_{0}\cdots \tilde{Y}_{\ell-1})=0,
\end{equation}
Combined with \eqref{eq:sum}, \eqref{eq:summand1}, and the nonnegativity of mutual information, this proves  \eqref{eq:MarkovK_strong_conv} for $\ell\in\{2,\ldots, k\}$.

\section{Conclusions and Outlook}
We derived the strong converse result for testing against independence over a $K$-hop network with $K$ decision centers and under a Markov chain assumption regarding the source sequences observed at the  terminals. Our strong converse proof is based on a change of measure argument similar to Gu and Effros \cite{GuEffros_1,GuEffros_2} and to Tyagi and Watanabe \cite{tyagi2019strong}. However, to obtain the desired Markov chain, we did not rely on the variational characterization of the weak converse result, as suggested by Tyagi and Watanabe \cite{tyagi2019strong}, nor did we use the blowing-up lemma or hypercontractivity arguments as in the proof for $K=2$  \cite{Vincent}. Instead, an easier proof is proposed that relies on showing the validity of the Markov chains in the limit of infinite blocklengths.  Our method can also be used for related scenarios, for example to establish the well-known strong converse for the Wyner-Ziv source coding problem.

\bibliographystyle{ieeetr}
\bibliography{references}
\end{document}